\documentclass[iop]{emulateapj}

\usepackage{epsfig}
\usepackage{color}
\usepackage{natbib}
\usepackage[normalem]{ulem}

\def\simgt{\lower.5ex\hbox{$\; \buildrel > \over \sim \;$}}
\def\simlt{\lower.5ex\hbox{$\; \buildrel < \over \sim \;$}}

\def\hii{\hbox{H{\small II}}}

\def\amin{\ifmmode^{\prime}\else$^{\prime}$\fi}
\def\asec{\ifmmode^{\prime\prime}\else$^{\prime\prime}$\fi}

\def\simgt{\lower.5ex\hbox{$\; \buildrel > \over \sim \;$}}
\def\simlt{\lower.5ex\hbox{$\; \buildrel < \over \sim \;$}}

\newcommand\rosat{{\it ROSAT\/}}
\newcommand\chandra{{\it Chandra}}

\newcommand\xmm{{\it XMM-Newton}}

\newcommand\swift{{\it Swift\/}}
\newcommand\nustar{{\it NuSTAR\/}}

\newcommand\fermi{{\it Fermi\/}}

\newcommand\degree{ .\!^{\circ} }
\def\sh2{\hbox{{\sl Sh}~2--104}}
\def\mgo{MGRO~J2019$+$37}
\def\ver{VER~J2019$+$368}
\def\nuneb{NuSTAR J201744.3$+$364812}
\def\xmmu{3XMM J201744.7$+$365045}
\def\fgl{3FGL~J2017.9$+$3627}
\def\psr{PSR~J2021$+$3651}

\slugcomment{Version of 2016 Apr 15}
\slugcomment{To be Submitted to The Astrophysical Journal}

\shorttitle{Hard X-ray Emission from \sh2}
\shortauthors{Gotthelf et al.}

\begin{document}

\title{ Hard X-ray Emission from \sh2: A NuSTAR search for Gamma-ray Counterparts}




\author{
E. V. Gotthelf\altaffilmark{1,2},
K. Mori\altaffilmark{1},
E. Aliu\altaffilmark{2},
J. M. Paredes\altaffilmark{2},
J. A. Tomsick\altaffilmark{3}, 
S. E. Boggs\altaffilmark{3},
F. E. Christensen\altaffilmark{4}, 
W. W. Craig\altaffilmark{3,5}, 
C. J. Hailey\altaffilmark{1}, 
F. A. Harrison\altaffilmark{6},
J. S. Hong\altaffilmark{7},
F. Rahoui\altaffilmark{3,11},
D. Stern\altaffilmark{9}, 
W. W. Zhang\altaffilmark{10}
}

\altaffiltext{1}{Columbia Astrophysics Laboratory, Columbia University, 550 West 120th Street, New York, NY 10027-6601, USA; eric@astro.columbia.edu}
\altaffiltext{2}{Departament de F\'{\i}sica Qu\`antica i Astrof\'{\i}sica, Institut de Ci\`encies del Cosmos, Universitat de Barcelona, IEEC-UB, Mart\'{\i}\ i Franqu\`es 1, 08028, Barcelona, Spain}
\altaffiltext{3}{Space Sciences Laboratory, University of California, Berkeley, CA 94720, USA}
\altaffiltext{4}{DTU Space-National Space Institute, Technical University of Denmark, Elektrovej 327, 2800 Lyngby, Denmark}
\altaffiltext{5}{Lawrence Livermore National Laboratory, Livermore, CA 94550, USA}
\altaffiltext{6}{Cahill Center for Astronomy and Astrophysics, California Institute of Technology, Pasadena, CA 91125, USA}
\altaffiltext{7}{Harvard-Smithsonian Center for Astrophysics, Cambridge, MA 02138, USA}
\altaffiltext{8}{European Southern Observatory, Karl Schwarzchild-Strasse 2, 85748 Garching bei M{\"u}nchen, Germany}
\altaffiltext{9}{Jet Propulsion Laboratory, California Institute of Technology, 4800 Oak Grove Drive, Pasadena, CA 91109, USA}
\altaffiltext{10}{NASA Goddard Space Flight Center, Greenbelt, MD 20771, USA}
\altaffiltext{11}{Department of Astronomy, Harvard University, 60 Garden Street, Cambridge, MA 02138, USA}
\begin{abstract}

  We present \nustar\ hard X-ray observations of \sh2, a compact \hii\
  region containing several young massive stellar clusters (YMSCs).
  We have detected distinct hard X-ray sources coincident with
  localized VERITAS TeV emission recently resolved from the giant
  gamma-ray complex \mgo\ in the Cygnus region.  Faint, diffuse X-ray
  emission coincident with the eastern YMSC in \sh2\ is likely the
  result of colliding winds of component stars. Just outside the radio
  shell of \sh2\ lies \xmmu\ and a nearby nebula \nuneb, whose
  properties are most consistent with extragalactic objects. The
  combined \xmm\ and \nustar\ spectrum of \xmmu\ is well-fit to an
  absorbed power-law model with $N_{\rm H} =
  (3.1\pm1.0)\times10^{22}$~cm$^{-2}$ and photon index $\Gamma =
  2.1\pm0.1$. Based on possible long-term flux variation and the lack
  of detected pulsations ($\le43\%$ modulation), this object is likely
  a background AGN rather than a Galactic pulsar. The spectrum of the
  \nustar\ nebula shows evidence of an emission line at $E=5.6$~keV
  suggesting an optically obscured galaxy cluster at $z=0.19\pm0.02$
  ($d = 800$~Mpc) and $L_X = 1.2\times10^{44}$~erg\,s$^{-1}$.
  Follow-up \chandra\ observations of \sh2\ will help identify the
  nature of the X-ray sources and their relation to \mgo.  { We
    also show that the putative VERITAS gamma-ray excess south of \sh2\ is
    most likely associated with the newly discovered 
    \fermi\ pulsar PSR~J2017$+$3625 and not the \hii\ region.}


\end{abstract}


\section{Introduction}

\begin{figure}[!t]
\centerline{
\hfill
\psfig{figure=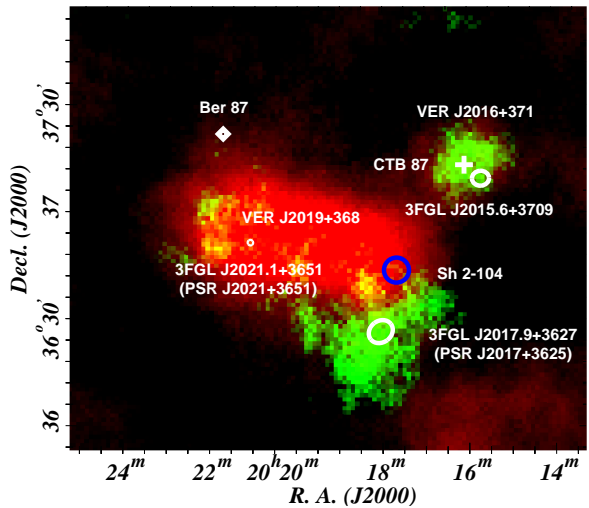,width=0.90\linewidth,angle=0}
\hfill
}
\caption{\small 
The TeV gamma-ray map of the \mgo\ region resolved into distinct
sources using VERITAS observations.  Note that the Milagro source
itself fills the field-of-view.  Superimposed are VERITAS images in
two energy bands, $0.6-1$~TeV (green) and $> 1$~TeV (red), that clearly
separate the Milagro emission into at least three distinct emission
components, each associated with a \fermi\ source ({\it white circles}). 
The harder, extended TeV emission, \ver, is associated with the pulsar
PSR~J2021$+$3651; \sh2, denoted by the blue circle, may contribute
a western component of this emission.  Modified from \cite{Aliu2014}.}
\label{fig:tev_image}
\end{figure}

\mgo\ is the brightest Milagro gamma-ray source in the Cygnus region,
with $80\%$ of the Crab Nebula flux at 20 TeV \citep{Abdo2007}.  The
origin and nature of \mgo\ has long been the subject of debate as its
$\sim 1^{\circ}$ extent overlaps several supernova remnants (SNRs),
\hii\ regions, Wolf-Rayet stars, $>100$ MeV gamma-ray sources, {one or
more \fermi\ pulsars} and a hard X-ray transient.  Recent TeV
observations on $\sim 6^{\prime}$ scales using the VERITAS telescope
clearly resolve out the giant gamma-ray complex into at least three
distinct TeV emission regions, each coincident with a \fermi\ source
\citep{Aliu2014}. The bulk of the VERITAS emission from \mgo\ falls
into the elongated ($1\fdg1\times 0\fdg6$), spectrally distinct
(harder) source \ver\ (see Figure~\ref{fig:tev_image}).

\citet{Paredes2009} argues that the \fermi\ pulsar PSR~J2021$+$3651
\citep{Roberts2002} at the eastern edge of \ver\ is not sufficiently
energetic to power all the gamma-ray flux in the region, based on the
time required for electrons to diffuse and fill the large emitting
volume relative to their cooling lifetime. Instead, these authors
suggest that massive star-forming activity associated with the \hii\
region Sharpless~104 (herein \sh2) can contribute to the gamma-ray
flux from \ver, possibly through wind collisions or interactions of
protostar jets with the surrounding medium \citep{Torres2004}.

The well-studied \sh2\ lies beyond the Cygnus Galactic arm,
$4.0\pm0.5$~kpc away, and contains at least two ultra compact massive
stellar clusters within its $\sim 7'$ radio diameter 
\citep{Paredes2009}.  The massive CO clouds around the star clusters
suggests \sh2\ as a prototype of massive-star formation triggered by
the expansion of an \hii\ region \citep{Deharveng2003}. An associated
H-alpha nebula is clearly resolved in the DSS POSSII-J image, likely
powered by a central O6 V star ionizing the region
\citep{Lahulla1985}, and possibly a bright nearby IRAS source.

A serendipitous \xmm\ observation of the \mgo\ field caught the
eastern half of the \sh2\ radio shell at the edge of the
field-of-view. Image analysis by \citet{Zabalza2010} revealed faint
emission in this short exposure (20~ks), just above the noise level,
that suggested several point sources within the radio shell. Most
notably, these include ones overlapping the central star, coincident
with a $2\sigma$ \rosat\ source, and the eastern YMSC.  Just outside
the radio shell lies \xmmu\ and a barely detected nebula $\sim
2^{\prime}$ in diameter.  These results open the possibility of
identifying a low energy counterpart to the gamma-ray emission, and
help identify its origin.

As part of the \nustar\ Galactic Survey program we have obtained broad
band X-ray observations of \sh2. In this paper, we report the
detection of hard X-ray emission from the eastern YMSC, and from the
\xmm\ source and the nearby diffuse nebula. We consider the possibility that
these sources are related to the star formation regions and/or
associated with gamma-ray emission. Alternatively, the latter two
sources may have an unrelated extragalactic origin.

\section{\nustar\ observations of \sh2} 

\nustar\ observed the \hii\ region \sh2\ on 2014 Oct 21 (ObsId \#30001048)
followed by a second overlapping observation, offset to the south,  on 
2014 Nov 13 (ObsId \#30001049).  
\nustar\ consists of two co-aligned X-ray telescopes,
with corresponding focal plane modules FPMA and FPMB that provide
$18^{\prime\prime}$ FWHM imaging resolution over a 3--79~keV X-ray
band, with a characteristic spectral resolution of 400 eV FWHM at 10
keV \citep{Harrison2013}.  The reconstructed \nustar\ coordinates are
accurate to $7\farcs5$ at the 90\% confidence level. The relative timing
accuracy of \nustar\ is $\sim 2$~ms rms, after correcting for thermal
drift of the on-board clock, with the absolute timescale shown to be
better than $< 3$~ms \citep{Mori2014, Madsen2015}.

The data were processed and analyzed using {\tt FTOOLS}
24Jan2014\_V6.15.1 ({\tt NUSTARDAS} 09Dec13\_v1.3.1) with \nustar\
Calibration Database (CALDB) files of 2013 August 30.  The resulting
data set provides a total of 80.5~ks and 91.6~ks of net good time for
the two pointings, respectively, after removing intervals of high
background rates. We also exclude a bright arc of stay light that
contaminates the eastern edge of the field-of-view in both detectors
during the first observation. The extracted spectra combined data from
both FPM detectors, grouped into appropriate spectral fitting
channels, and modeled using the {\tt XSPEC} (v12.8.2) spectral fitting
package (Arnaud 1996). All spectral fits use the {\tt TBabs}
absorption model in {\tt XSPEC} with the {\tt wilm} Solar abundances
\citep{Wilms2000} and the {\tt vern} photoionization cross-section
\citep{Verner1996}.

\subsection{Image analysis}

\begin{figure*}[!t]
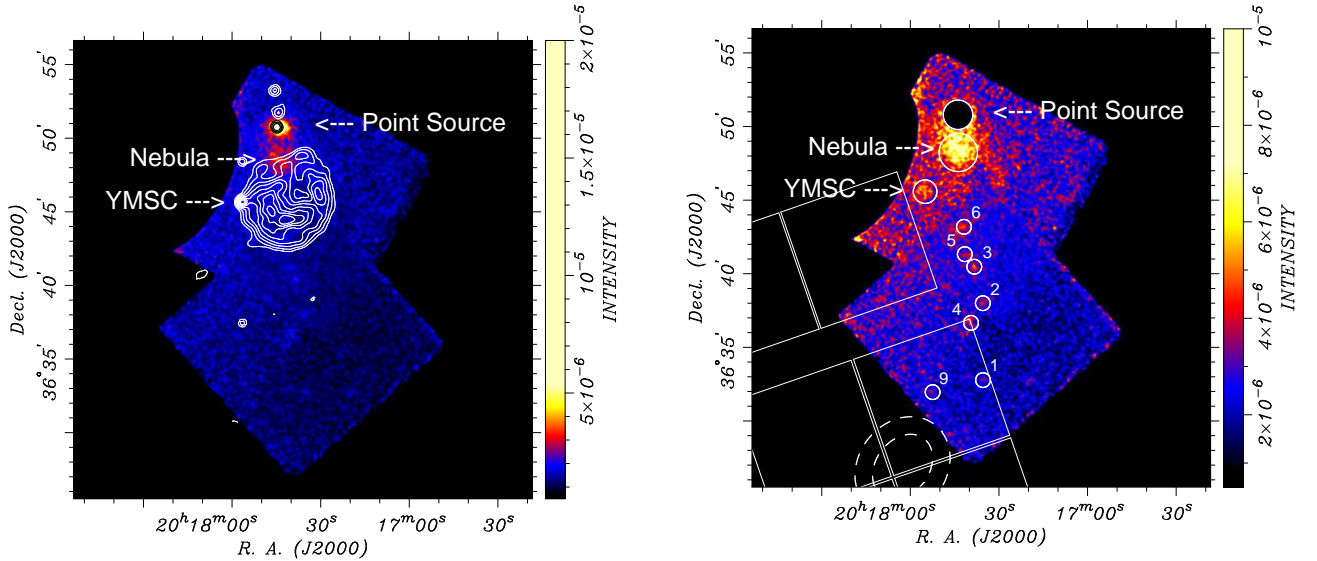

\hfill
\psfig{figure=sh2-104_2fgl_mosaic_35_460_corr_cut_sm1.5.ps,height=0.45\linewidth,angle=-90}
\hfill
\psfig{figure=sh2-104_2fgl_mosaic_35_460_corr_diffuse_sm1.5.ps,height=0.45\linewidth,angle=-90}
\hfill
\caption{\small  \nustar\ exposure-corrected and smoothed 3-79~keV 
  X-ray images of the field containing \sh2. 
  {\it Left ---} A bright \nustar\ point source is evident up to 30~keV
  and significant X-ray nebula is found in the $3-20$~keV band just
  south of the point source.  Both of these sources are of unknown
  origin.  The nebula overlaps the rim of the \hii\ region \sh2, here
  outlined by the GMRT 610~MHz radio contours of \cite{Paredes2009}. Faint hard X-ray
  emission is also seen at the location of the eastern young massive
  stellar cluster (YMSC) enclosed by the radio spur.  A bright arc on
  the edge of the image, due to stray light hitting the focal plane,
  has been excised. {\it Right ---} The same image with the bright point 
  source removed and scaled to highlight diffuse emission. Indicated 
  ({\it circles}) are the detected \nustar\ sources listed in 
   Table~\ref{tab:sources}. Also shown is the location of the nearby 
  \chandra\ observation ({\it outline}) of \fgl\ ({\it contours}).}
\label{fig:images}
\end{figure*}

Figure~\ref{fig:images} presents the exposure-corrected 3--79~keV
\nustar\ images of the \sh2\ field, combining data from both FPM
detectors. The images are smoothed using a
$\sigma=3.\!^{\prime\prime}7$ Gaussian kernel and scaled linearly.
Most prominently, we detect a hard ($> 30$~keV) point source just
north of the \sh2\ complex and a poorly resolved nebula $2.3^{\prime}$
below it, roughly $2.4^{\prime}$ in diameter. These sources clearly
correspond to faint X-ray emission seen in a short, 20~ks, 2007 \xmm\
observation, detected serendipitously, at the very edge of the
field-of-view \citep{Zabalza2010}.

The two bright \nustar\ sources are embedded in enhanced diffuse
emission that overlaps, at least in part, with the \sh2\ radio
nebula. Of particular interest is a faint nebula detected at the radio
spur of \sh2, coincident with the eastern YMSC discussed in
\cite{Paredes2009}.  Its spatial extent of $r \sim 0\farcm5$,
corrected for the PSF, is consistent with the size of the optical
cluster. Further south we find evidence of several other faint
sources, including a \swift\ X-ray source obtained as part of the
follow-up program of \fermi\ sources. Table~\ref{tab:sources} presents
the list of detected \nustar\ sources along with the significance of
detection computed by {\tt wavdetect}. The source coordinates are
accurate to $\approx 4\farcs0$, registered using \xmmu, the counterpart to the
bright \nustar\ point source.  Finally, we note that no hard X-ray
counterpart is detected for the \rosat\ source at the center of \sh2\
\citep{Paredes2009}. This suggest a soft X-ray source, likely thermal
emission from the bright star in the central cluster, or a
low-temperature colliding wind binary within this cluster.

%
%

\begin{deluxetable*}{cccccl}
\tablecaption{\nustar\ Sources in the \sh2\ Field} 
\tablecolumns{6}
\tablehead{ \colhead{\#} & \colhead{ R.A. }& \colhead{ Decl. } & \colhead{Net}  & \colhead{Sig.} & \colhead{Comment} \\
\colhead{ }   & \colhead{ } & \colhead{ }  & \colhead{Counts}  & \colhead{(Sigma) }   & \colhead{ } }
\startdata
1  & 20 17 35.97  & +36 32 45.0 &  $  61 \pm 14$ &  4.4 &\\
2  & 20 17 36.02  & +36 37 58.5 &  $  73 \pm 16$ &  5.2 &\swift\ source  \\
3  & 20 17 38.91  & +36 40 27.1 &  $  45 \pm 13$ &  3.7 &\\
4  & 20 17 40.02  & +36 36 38.3 &  $  24 \pm  8$ &  3.1 &\\
5  & 20 17 42.13  & +36 41 18.2 &  $  50 \pm 13$ &  4.0 &\\
6  & 20 17 42.41  & +36 43 09.7 &  $  48 \pm 13$ &  4.1 &\\  
7  & 20 17 44.32  & +36 48 12.9 &  $2219 \pm 82$ &   27.0\phantom{0} &  Nebula \nuneb  \\
8  & 20 17 44.42  & +36 50 46.4 &  $1406 \pm 54$ &   35.0\phantom{0} & \xmmu\ \\
9  & 20 17 53.02  & +36 31 56.0 &  $  52 \pm 12$ &  4.9 &\\
10 & 20 17 55.57  & +36 45 33.1 &  $ 197 \pm 34$ &  5.8 &\sh2\ stellar cluster / H{\tiny II} region
\enddata
\tablecomments{Coordinate system is registered to $\approx 4\farcs0$ accuracy
  using \xmmu, the counterpart to the bright \nustar\ point source.
}
\label{tab:sources}
\end{deluxetable*}

\subsection{Spectral analysis}

To preface our spectral analysis we note that the \nustar\ low-energy
response ($3$~keV) is too hard to constrain the absorbing column for
a typical source with $N_{\rm H} < 10^{22}$~cm$^{-2}$. In the following spectral
fits using \nustar\ data alone, for definitiveness, we hold the column
density fixed to a fiducial value of the Galactic total. Generally,
the range of likely column density here is found to have no
significant effect on the resulting spectral parameters.  We include
both a neutral
Hydrogen\footnote{\url{http://heasarc.gsfc.nasa.gov/cgi-bin/Tools/w3nh/w3nh.pl}}
and a molecular Hydrogen component to the column density, to take into
account significant local CO emission \citep[see][]{Dame2001}. We
compute a total Galactic column density of $N_{\rm H} = n_{\rm
  H{\small I}} + 2n_{H_2} = 2.4\times10^{22}$~cm$^{-2}$, a value
consistent with the results of the combined \xmm\ and
\nustar\ fit to the spectrum of \xmmu, presented below. In all cases,
the quoted spectral uncertainties are at the 90\% confidence level for one or
two interesting parameter(s) for the one and two component spectral
fits, respectively.

The high-energy emission from the eastern YMSC is of great interest,
as this source is a natural candidate for the observed gamma-ray emission.  We
extracted a \nustar\ spectrum from YMSC using a $r < 0.8^{\prime}$
aperture in the usable 3--10~keV range. This yields a total of 680
counts of which 71\% are from background contamination, as estimated
from counts extracted from an adjacent aperture ($r < 1.4^{\prime}$) on same chip
of each FPM. We consider several appropriate spectral models as the
quality of the spectrum is not sufficient to distinguish between
then. Under the assumption that the X-ray emission is due to colliding
winds of component stars in the cluster we fit the {\tt raymond}
thermal plasma model in {\tt XSPEC} \citep[][and
updates]{Raymond1977}.  The best-fit temperature is $kT =
1.2-3.5$~keV with a $\chi^2 =1.1$ for 19 degrees-of-freedom (DoF). The
2--10~keV unabsorbed flux is $1.7\times
10^{-13}$~erg~cm$^{-2}$~s$^{-1}$. We estimate a source luminosity of
$L \sim 10^{33}$~erg~s$^{-1}$ from the plasma cooling curve
\citep[e.g.,][]{Maio2007} and the derived emission measure, computed
from the model normalization and a distance of 4~kpc to \sh2.  
%
%
We also consider a non-thermal model that can result from accelerated
particles of past supernovae and/or quiescent or faint X-ray
binaries. For a simple power-law model, the photon index is $\Gamma =
3.5(2.4-4.7)$ with a similar $\chi^2$ and flux as found for the
thermal model.



For \xmmu, we extracted a high quality \nustar\ spectrum using a
$r=45^{\prime\prime}$ source aperture and a $r=1.4'$ background region
offset from the source.  The source spectrum is found to dominate the
background up to 20 keV, but emission is evident to at least 30~keV.
The spectrum is well-fitted in the 3--20~keV energy band to an
absorbed power-law model with the column density held fixed to the
Galactic total. The best fit spectral index is $\Gamma = 2.0 \pm 0.1$
with  $\chi^2 = 0.82$  for 28 DoF. The
unabsorbed flux in the 2--10~keV band is $(1.4\pm0.1)\times
10^{-12}$~erg~cm$^{-2}$~s$^{-1}$. A blackbody model is excluded by the
fit, as is, for lack of line features, thermal plasma models.
The spectral results are presented Table~\ref{tab:specfit}.

\begin{figure}[]
\centerline{
\hfill
\psfig{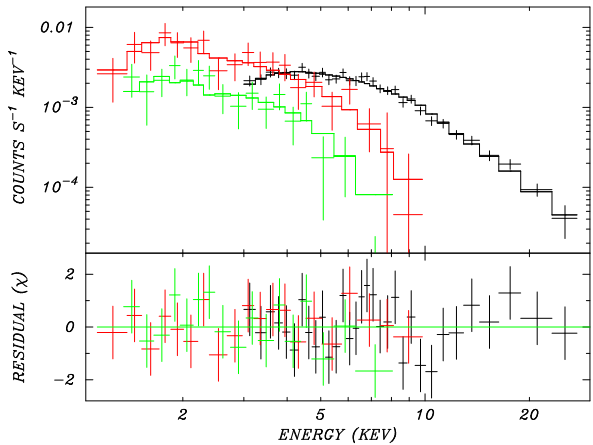}
\hfill
}
\caption{\nustar\ and \xmm\ spectra of \xmmu. The spectra are fitted
  simultaneously to an absorbed power-law model with their
  normalizations left free.  The upper panel present \nustar\ (black)
  and \xmm\ EPIC~pn (red) and EPIC~MOS (green) spectral histograms
  along with the best-fit model (solid lines) given in
  Table~\ref{tab:specfit}. The lower panel shows the residuals from
  the best-fit model in units of sigma.}
\label{fig:ps}
\end{figure}

\begin{figure*}[]
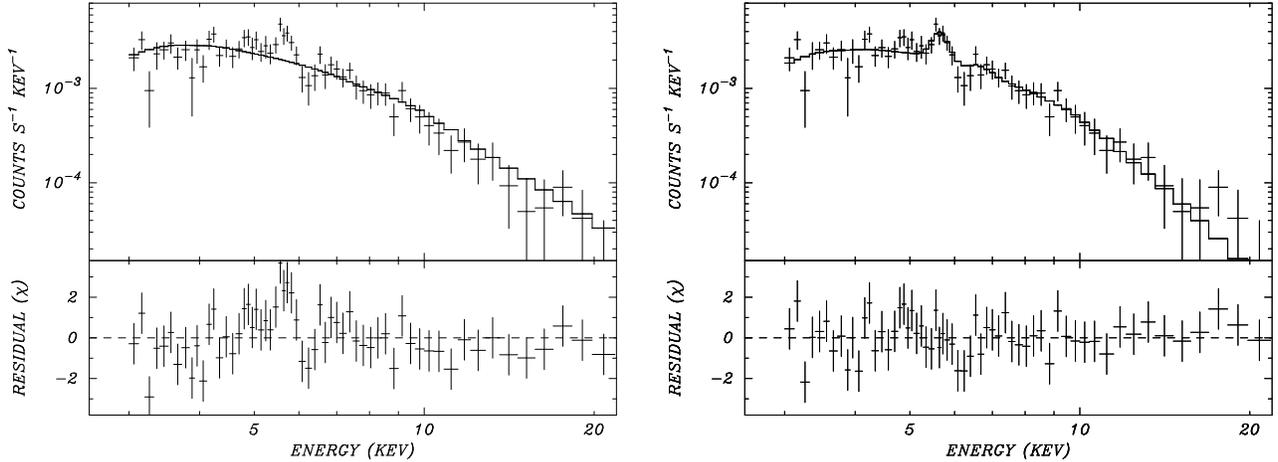

\vskip 0.1in
\centerline{
\hfill
\psfig{figure=sh2_104_nustar_nebula_flt_pl.ps,height=0.45\linewidth,angle=-90}
\hfill
\psfig{figure=sh2_104_nustar_nebula_flt_ray.ps,height=0.45\linewidth,angle=-90}
\hfill
}
\caption{\nustar\ spectra of the new X-ray nebula \nuneb.
  {\it Left ---} The spectrum fitted with an absorbed power-law (solid
  lines).  The lower panel shows the residuals from the best-fit model given in the text, in units of sigma.
  {\it Right ---} The same spectrum fitted with the
  addition of an emission line.}
\label{fig:nebula}
\end{figure*}

\begin{deluxetable*}{lccc}[b]
\tablecaption{\nustar\ and \xmm\ Spectra of \xmmu} 
\tablecolumns{4}
\tablehead{ \colhead{Parameter}   &  \colhead{\nustar\ only}  &  \colhead{\xmm\ only} &  \colhead{\nustar\ + \xmm\ \tablenotemark{a}}  }
\startdata
$N_{\rm H}$ (cm$^{-2}$)    & $2.4 \times 10^{22}$ (fixed) & $3.4(1.8-5.6) \times 10^{22}$& $(3.1\pm 1.0) \times 10^{22}$        \\
$\Gamma$                   & $2.0\pm0.1$                  & $1.4-3.1$               & $2.1\pm 0.1$                  \\
$F_{\rm abs}$ (2--10 keV)  & $1.2\pm0.1 \times 10^{-12}$  &   $3.5 \times 10^{-13}$ & $1.2\pm0.1 \times 10^{-12}$       \\
$F_{\rm unabs}$ (2--10 keV)& $1.3 \times 10^{-12}$        &   $4.6 \times 10^{-13}$ & $1.4 \times 10^{-12}$       \\
$\chi^2_{\nu}$ (dof)       & 0.82 (29)                    & 0.68 (36)               & 0.70 (62)                                
\enddata
\tablecomments{Power-law model fits are obtained in the 0.5--10~keV
  and 3--20~keV energy bands for the \xmm\ and \nustar\ spectra,
  respectively.  For the joint \nustar\ and \xmm\ spectral fits, the
  indexes and column densities are linked. The uncertainties are 90\%
  confidence limits for two interesting parameters, except for the
  \nustar\ only data, which is for one interesting parameter.
  The given fluxes are in units of erg~cm$^{-2}$~s$^{-1}$. 
 }
\tablenotetext{a}{The listed flux values are for the \nustar\ spectra, while the \xmm\ spectra are jointly fit with a relative normalization factor to account for the flux variation.}
\label{tab:specfit}
\end{deluxetable*}

To better estimate the source column density for \xmmu\ we extracted
and fit the \xmm\ spectrum simultaneous with the \nustar\ data,
allowing the flux normalization to be independent.  This resulted in
$N_{\rm H} = (3.1 \pm 1.0)\times 10^{22}$~cm$^{-2}$ and spectral index
$\Gamma = 2.1 \pm 0.1$ (see Figure~\ref{fig:ps}). We note that the
measured column is consistent with our estimate of the Galactic total.
Compared to the \xmm\ flux measurements obtained 7 years earlier, the
\nustar\ value is lower, formally by a factor of $\sim 3$. However,
the former flux is not well established due to poor statistics, the
high background, and the far off-axis source detection on the edge of
the \xmm\ EPIC instruments, and the relative flux calibration between instruments.

To examine the X-ray nebula \nuneb, we extracted spectra
using a $r=1.4^{\prime}$ radius aperture and the background region
defined above.  This aperture encompassed nearly all the nebula extent
to the background level.  Of the $\sim 3400$ aperture counts in the
3--20~keV optimal energy band, $\sim 1400$ (42\%) are attributed to
the background.  A fit to the nebula spectrum using an absorbed
power-law model in the 3--20~keV band with the total Galactic column
density yields a photon index $\Gamma = 2.5-2.8$ with a $\chi^2_{\nu}
= 1.6$ for 57 DoF. The 2--10~keV absorbed flux is $(6.4-7.5)\times
10^{-13}$~erg~cm$^{-2}$~s$^{-1}$ (90\% confidence level) and
the unabsorbed flux is $8.7 \times 10^{-13}$~erg~cm$^{-2}$~s$^{-1}$.

The poor $\chi^2$ statistic for this fit is mainly due to a line-like feature
around $5.6$~keV (see Figure~\ref{fig:nebula}).  No similar feature
appears in the background or the \nustar\ spectrum of the point source
\xmmu.  Introducing a Gaussian line to the fit, to better characterize
the continuum, yield a line measured at $5.6\pm 0.2$~keV and results
in an excellent fit statistic of $\chi^2_{\nu} = 0.94$ for 54 DoF.
The F-test value of $13.09$ corresponds to a false positive
significance of the added spectral line of $\wp = 2.3\times 10^{-5}$.
However, this significance should be interpreted with care
\citep{Protassov2002}. Formally, an analysis of the F-test probability
using the {\tt XSPEC} script {\tt simftest} confirms a highly
significant detection of an emission line feature associated with the
source.

As no emission line is known at this energy we consider the
possibility of a redshifted Fe line from a galaxy cluster hidden
behind the Galactic plane. A spectral fit using the {Raymond-Smith}
model for a thermal plasma produced an excellent fit again, with
$\chi^2_{\nu} = 0.86$ for 55 DoF, with the column density fixed at the
Galactic total. The best-fit parameters give a
$kT=5.6_{-0.83}^{+1.3}$~keV, redshift $z=0.19\pm0.02$, and abundance
$0.66 Z_{\sun}$. The 2--10~keV absorbed flux is
$6.4_{-0.6}^{+0.3}\times 10^{-13}$~erg~cm$^{-2}$~s$^{-1}$ (90\%
confidence level) and the unabsorbed flux is $7.5 \times
10^{-13}$~erg~cm$^{-2}$~s$^{-1}$. We note that the measured range of
$kT$ is essentially independent of the column density, from zero to
$3.1\times10^{22}$~cm$^{-2}$, the point source value.  This $N_{\rm
H}$ range yields a measured temperature range of $kT = 5.3-6.4$~keV,
comparable to the uncertainty in $kT$ obtained using the Galactic
column.


For a galaxy cluster at the implied redshift distance of 800~Mpc, the
total X-ray luminosity is $L_x = 1.2\times10^{44}$~erg~s$^{-1}$.  This
is within an order of magnitude of the value derived from the
luminosity - temperature relation for clusters
\citep{Novicki2002}. Moreover, based on the inferred temperature, the
observed nebula size is well predicted for a putative cluster
\citep{Mohr2000}.  Given the large uncertainties in these relations we
take the results as reasonably evidence that the X-ray nebula is due to
a background galaxy cluster unrelated to the gamma-ray emission.

\subsection{Timing Analysis}

The high time resolution of \nustar\ allows a search for pulsations
from \xmmu\ down to periods of $P \sim 4$~ms, covering the expected
range for a rotation-powered pulsar.  For a timing analysis, photon
arrival times were converted to barycentric dynamical time (TDB) using
the \xmm\ coordinates.  The \nustar\ light curve is found to be stable
during the observation on all timescales. A Fast Fourier Transform
(FFT) finds no evidence of red noise, indicative of accreting systems
in the power spectrum. We also searched for a coherent signal using
both the FFT method and the $Z^2_n$ test statistic for $n =1,2,3, 5$,
and the H-test, to be sensitive to both broad and narrow pulse
profiles. We initially restricted the timing search to photon energies
in the 3--25 keV range and used an aperture of $30^{\prime\prime}$ to
optimize the signal-to-noise ratio.  We repeated our search for an
additional combination of energy ranges $3<E<10$~keV, $10<E<25$ keV
and aperture sizes $r> 10^{\prime\prime}$.  None of these resulted in
a significant detection.  After taking into account the estimated
background emission, we place an upper limit on the pulse fraction
$f_p\le43\%$ for a sinusoidal signal in the 3--25 keV band for the
$20^{\prime\prime}$ aperture.

\section{Discussion}

TeV gamma-ray emission from \mgo\ is well-separated into three regions
by high-resolution VERITAS observations, each associated with a
\fermi\ source \citep{Aliu2014}.  To the north, VER~J2016$+$371 is
likely associated with \fermi\ emission from the blazar B2013$+$370
\citep{Kara2012} and/or the filled-center supernova remnant CTB~87
\citep{Aliu2014}. { To the south,  localized VERITAS emission,
  significant at the $3\sigma$ confidence level, is coincident with
  the recently discovered \fermi\ pulsar, PSR~J2017$+$3627, discussed
  below.}
To the east, the large elliptical morphology of the spectrally harder
\ver\ suggests a blend of two overlapping sources.  If the arguments
of \citet{Paredes2009} are correct, the \fermi\ pulsar \psr\ easily
accounts for the eastern most TeV emission from \ver, while \sh2\ may
be responsible for a western component. In the following, we use new
X-ray observations to explore possible origins for an eastern
component of \ver.


The \nustar\ data reveals faint emission from the eastern YMSC of
\sh2, with an inferred luminosity of $L \sim 10^{33}$~erg~s$^{-1}$. 
In other YMSCs, hard $>$3\,keV X-rays have been seen at similar
levels from either point sources \citep{clark08} or diffuse 
emission \citep{townsley11}.  For the point source case, although
it is possible for hard X-rays to be produced by an accreting
compact object, a small number of isolated OB supergiants or
Wolf-Rayet stars, colliding wind binaries (CWBs), or possibly 
even a single CWB could produce hard X-ray emission at the levels 
we observe \citep{clark08,sugawara15}.  If a thermal model is the 
correct characterization of the {\em NuSTAR} spectrum, then the 
plasma temperature of $\sim$2--3\,keV would favor CWBs over
isolated massive stars \citep{bodaghee15,sugawara15}.  Diffuse
thermal or non-thermal X-ray emission could come from particles
accelerated by past supernova events or continuous acceleration
in CWB shocks \citep{muno06,townsley11}.  \cite{townsley11} 
report non-thermal diffuse emission from NGC~3576~N with a power-law 
spectrum with $\Gamma$$<$2.5.  The power-law fit to \sh2 YMSC
gives $\Gamma = 3.5^{+1.2}_{-1.1}$, which is just barely compatible
with NGC~3576~N.  

{ The coincidence of gamma-ray emission near star-forming regions
suggests a physical connection between the two -- for example, W49A
\citep{Brun2010}, Westerlund1 \citep{Luna2010}, and Carina Nebula,
although the gamma-rays from the latter is likely dominated by the
CWB $\eta$~Carinae \citep{tavani09,farnier11}.  However, this
connection remains far from clear, at least on an individual basis.
For the case of the YMSC in \sh2, we can consider several plausible
physical mechanisms for generating the associated gamma-rays. As
mentioned above, it is unlikely that the massive star binaries and
protostars in the YMSC are sufficiently energetic to power a
significant fraction of the TeV flux from the Milagro source, that is likely
hadronic in nature, if extended. However, the YMSC may contribute
gamma-rays to the western compact component of VER
J2019+368, via a leponic process.  This reduces the required
energy budget by about two orders of magnitude (see Paredes et al.
2009).}

{ The YMSC could also host other young and powerful non-thermal
sources associated with massive stars, such as high-mass
microquasars or massive binaries containing non-accreting pulsars
\citep[e.g.][]{Paredes2013,Dubus2013}.  These sources, potentially
hidden in hard X-rays by the dense environment in which they would
be embedded, could also contribute to the overall gamma-ray emission
from VER J2019+368.  Qualitatively, if only a fraction of the source
flux came from the YMSC, for exmaple a third, the energetic
requirements of the western component of VER J2019+368 would be
reduced, and a hadronic mechanism may be plausible.  The lack of
associated GeV emission from VER J2019+368 could be explained then
by hadronic models for which the lower energy emission is
suppressed, e.g., proton-proton interactions in the innermost region
of the winds of massive O and B stars, as suggested by \citet{Torres2004} 
\citep[see also][for scenarios in which the GeV emission is rather low 
as compared to the TeV emission]{Aharonian1996,Bosch-Ramon2005}.}


Alternatively, the substantial X-ray emission that lies outside the
radio nebula of \sh2\ perhaps signals a previously unidentified star
cluster responsible for the TeV emission.  However, no specific
optical or infrared counterpart is known. The coordinates of \xmmu\
are consistent to within $1\farcs1$ with the 20~cm arcsec radio point
source G74.840$+$0.660 \cite[FIRST Radio Survey;][]{White2005},
possibly the 327 MHz source WSRTGP~2015$+$3641 \citep{Taylor1996},
both nondescript radio objects in these Galactic surveys. A dedicated
radio observation of Cygnus at 610~MHz by \cite{Paredes2009} determined
a flux density of $33.46\pm0.08$~mJy for GMRT J201744.8$+$365045. Comparing
this to the 20~cm flux of 11.15~mJy \citep{White2005} yields a spectral
index of $\alpha = 1.2$, where $F_{\nu} = \nu^{-\alpha}$.

The combination of the radio and X-ray point source and diffuse
emission suggest a pulsar and its wind nebula, perhaps born in the
star formation region, which provides a natural source of seed photons
for generating upscattered gamma-rays \citep[cf.,
HESS~J1837--069/PSR~J1838--0655; ][]{Gotthelf2008}.  Although
the offset between the point source and the nebula is somewhat
unusual, PWN systems often show complex X-ray morphology, as
revealed by \chandra\ \citep[e.g., Crab, MSH~15$-$52; see][]{Kargaltsev2008}.

The X-ray spectrum of \xmmu\ is, however, somewhat steep for a pulsar,
more consistent with that of a hidden optical AGN behind the Galactic
plane. The radio spectrum also prefers an AGN interpretation over a
pulsar. The likely coincidence with a bright point-like radio source,
the lack of detected pulsations ($>43\%$), and the possible long term
variability, strengthens this interpretation. However it is worth
noting that the upper-limit on the modulation for an X-ray pulsar is
not strongly constraining.

The origin of the nebula \nuneb, and whether it is connected
to the point source, remains a mystery. The appearance of a possible
spectral line at an unexpected energy suggests that this feature, if
astrophysical, is likely a red-shifted Fe line from a galaxy cluster
hidden behind the Galactic plane. The estimated luminosity and size of
the nebula, based on its implied redshift and temperature, is
consistent with this interpretation. The large column in the region
(e.g., $\sim 19$ magnitude of extinction in the V-band) could easily
account for the lack of an optical counterpart.

{ Finally, it is possible that the TeV gamma-ray emission near
  \sh2\ might be associated with the \fermi\ source \fgl, $0\degree3$
  to the south of the \hii\ region. A search for pulsations from \fgl\
  by the Einstein@Home distributed computing pulsar project
  \citep{Anderson2006,Allen2013} detected a 167~ms signal, consistent with a
  2~Myr old rotation-powered
  pulsar\footnote{https://einstein.phys.uwm.edu/gammaraypulsar/
    FGRP1\_discoveries.html} (Clark, C., in prep.). The inferred spin-down power of
  PSR~J2017+3625, $\dot E = 1.2\times10^{34}$~erg~s$^{-1}$, suggests
  that it likely lies at a distance of 450~pc, given its gamma-ray
  flux of $4.8\times10^{-11}$~erg~s$^{-1}$~cm$^{-2}$ \citep{Acero2015}
  and a gamma-ray efficiency of $L_{\rm GeV} / \dot E = 0.1$, typical
  for a \fermi\ pulsar. On the other hand, the lack of significant
  X-ray detection of a candidate NS in the unpublished 10~ks \chandra\
  observation (ObsID 14699) suggests that the pulsar is further away.
  For this observation, we estimate a flux limit of $F_x(2-10\ {\rm
    keV}) \lesssim 2 \times 10^{-14}$~erg~s$^{-1}$~cm$^{-2}$ for a
  typical pulsar power-law spectrum ($N_{\rm H} =
  1.0\times10^{22}$~cm$^{-2}$; $\Gamma = 1.5$). The predicted distance
  is then $\gtrsim 1$~kpc, based on the empirical relation between
  the X-ray luminosity of pulsars and their spin-down power
  \citep{Possenti2002}.  At this distance, the local TeV emission,
  estimated to be roughly $L(1-10 \ {\rm TeV}) \sim 3\times
  10^{-12}$~erg~s$^{-1}$, represents an efficiency of $L_{\rm TeV} /
  \dot E \sim 0.03$, plausible for a $>10^{5}$~yr pulsar \citep[e.g.,
  see][]{Kargaltsev2013}.}

{ In conclusion, it is possible that PSR~J2017$+$3625 accounts for
  most, if not all, of the coincident VERITAS TeV excess and that the
  harder TeV photons near \sh2\ remain unaccounted for. Our analysis
  of the X-ray data is inconclusive as to the connection between the
  \nustar\ sources, the \sh2\ region, and the overlapping gamma-ray
  emission.  Without further evidence, it is not yet possible to
  associate the Milagro gamma-ray emission with \sh2.  In this regard,
  it is important to determine the nature of the \nustar\ sources
  presented in this study. This will require high resolution \chandra\
  observations to allow a comparison between these sources and several
  overlapping optical/IR stars and an unclassified radio source, that
  may or may not be related to the X-ray and/or gamma-ray emission.}

\acknowledgements

This work was supported under NASA Contract No. NNG08FD60C, and made
use of data from the \nustar\ mission, a project led by the California
Institute of Technology, managed by the Jet Propulsion Laboratory, and
funded by the National Aeronautics and Space Administration. We thank
the \nustar\ Operations, Software and Calibration teams for support
with the execution and analysis of these observations. This research
has made use of the \nustar\ Data Analysis Software (NuSTARDAS)
jointly developed by the ASI Science Data Center (ASDC, Italy) and the
California Institute of Technology (USA). {  E.V.G. acknowledges partial
support by the National Aeronautics and Space Administration through
\xmm\ Award Number NNX15AG28G and Chandra Award Number G05-16061X,
issued by the Chandra X-ray Observatory Center, which is operated by
the Smithsonian Astrophysical Observatory for and on behalf of the
National Aeronautics Space Administration under contract NAS8-03060.
J.M.P. acknowledges support by the Spanish MINECO under grants
AYA2013-47447-C3-1-P, MDM-2014-0369 of ICCUB (Unidad de Excelencia
`Mar\'\i a de Maeztu'), and the Catalan DEC grant 2014 SGR~86 and
ICREA Academia.}

\bibliography{sh2-104_nustar_paper_v5}

\end{document}